# Proposal for a near-field optomechanical system with enhanced linear and quadratic coupling

**Hao-Kun Li[1]\*, Yong-Chun Liu[1], Xu Yi[1], Chang-Ling Zou[2], Xue-Xin Ren[1], and Yun-Feng Xiao[1]\*\***

[1]State Key Laboratory of Mesocopic Physics and School of Physics, Peking University, Beijing 100871, P. R. China

[2]Key Lab of Quantum Information (CAS), University of Science and Technology of China, Hefei 230026, P. R. China

\*Email address: hkli@pku.edu.cn

\*\*URL: www.phy.pku.edu.cn/~yfxiao/index.html

We propose a realistic system with separated optical and mechanical degrees of freedom, in which a high-mechanical-quality silicon nitride membrane is placed upon a high-optical-quality whispering gallery microcavity. The strongly enhanced linear optomechanical coupling, together with simultaneously low optical and mechanical losses in the present system, results in a remarkable single-photon cooperativity exceeding 300. This unprecedented cooperativity in the optomechanical system enables optical nonlinearity at low light intensities and holds great potential in generating, storing, and implementing quantum states. Moreover, the device gives rise to significantly stronger quadratic optomechanical coupling than achieved to date, which is desirable for measuring phonon shot noise with a high signal-to-noise ratio.

Optomechanical systems, in which macroscopic mechanical objects couple to electromagnetic degrees of freedom, have attracted growing interests over the past few years [1-4]. These systems not only offer great opportunities to generate and control quantum states of light and macroscopic mechanical oscillators, but also provide a promising arena for exploring the boundary between quantum and classical physics [5]. Experiments including quantum limited measurements [6], mechanical ground-state cooling [7, 8] and optomechanically induced transparency [9, 10] have made remarkable steps toward investigations and applications of quantum optomechanics. To realize optical nonlinearity and further quantum implementation such as generation, storage and control of quantum states in optomechanical systems, the effective coupling rate $G \equiv \sqrt{n}g$ must be strong enough to overcome the optical and mechanical damping ($\kappa$ and $\gamma_m$), and the environment heating ($\Gamma_{th} = n_{th}\gamma_m$), where $g$ represents the single-photon optomechanical coupling rate, $n$ stands for the mean number of drive-laser intracavity photons, $n_{th}$ denotes the mean thermal excitation number. Generally, the effective coupling can be enhanced by increasing the input optical power. Unfortunately, this inevitably causes the optical absorption-induced heating which seriously destroys quantum behaviors of optomechanical systems [11, 12]. With regard to this, a strong single-photon coupling together with low optical and mechanical losses is of primary concern in quantum applications of optomechanical systems.

Up to now, various optomechanical structures have been proposed, but they do not possess simultaneously large single-photon coupling rate, low mechanical and optical losses. For instance, optomechanical crystals [13, 14] with small optical mode volumes arrive at large coupling rate ($g/2\pi \sim 10^2$ kHz) but technically remain large $\gamma_m/2\pi$ (~ 0.1 MHz) and $\kappa/2\pi$ (~500 MHz); the membrane-in-the-middle setups [15] can achieve both small $\gamma_m/2\pi$ (~ 0.1 Hz) and $\kappa/2\pi$ (~ 0.1 MHz) by separating optical and mechanical degrees of freedom, but the coupling rate ($g/2\pi \sim 10$ Hz) is small due to the large cavity size; the side-coupling toroid-string structures [6, 16] also enable simultaneously low $\gamma_m/2\pi$ (~ 10 Hz) and $\kappa/2\pi$ (~ 5 MHz), while the coupling strength ($g/2\pi \sim 1$ kHz) is limited by insufficient interaction area between the optical field and the string.

In this paper, we propose a near-field optomechanical system in which a low $\gamma_m/2\pi$ (~ Hz) $Si_3N_4$ membrane [17] is placed upon a low $\kappa/2\pi$ (~ 5 MHz) silica microtoroid [18]. With this new design, the interaction area between the membrane and the optical field is significantly increased, which leads to over one-order-of-magnitude enhanced linear single-photon optomechanical coupling rate compared with the toroid-string structure, reaching $g/2\pi \sim 20$ kHz. As a result, the single-photon cooperativity (defined as $C = 4g^2/\kappa\gamma_m$) exceeds 300. This unprecedented cooperativity allows for optical nonlinearity at low light intensities. Furthermore, a large quadratic single-photon optomechanical coupling rate can be obtained, which ensures phonon shot noise detection with a high signal-to-noise ratio [19].

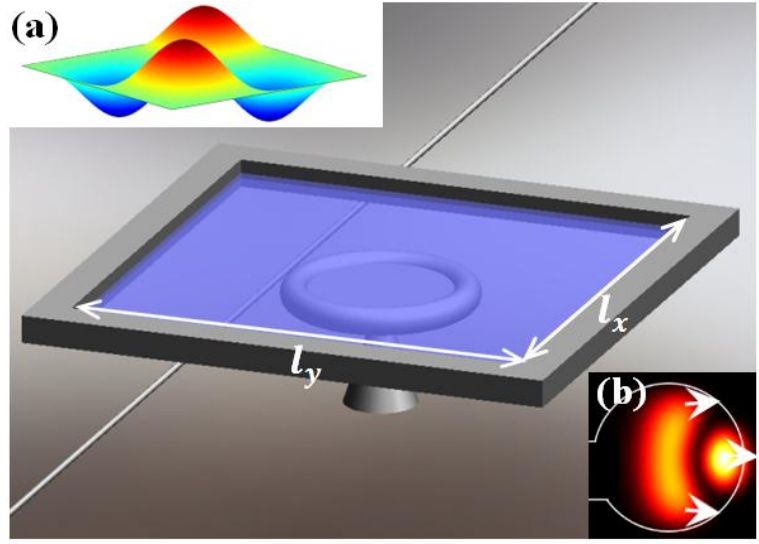

FIG. 1 (color online) Schematic illustration of the optomechanical system. Insets: (a) Illustration of membrane mechanical mode amplitude $u_{j,k}(x, y)$ with $(j, k) = (2, 2)$; (b) False-color representation of the squared transverse electric fields for the third-order TM optical whispering gallery mode in the silica microtoroid. White arrows denote the directions of the local electric fields.

Figure 1 shows the configuration of the present optomechanical system. The dimensions of the membrane are $l_x \times l_y \times h$, and the major and minor diameters of the microtoroid are $D$ and $d$. A fiber taper is used to couple light into and out of the microcavity. The membrane supports various mechanical modes with eigenfrequencies $\omega_m(j,k)/2\pi = \sqrt{T/4\rho}\sqrt{j^2/l_x^2 + k^2/l_y^2}$, where $j$ and $k$ represent mode indices, $\rho = 2.7$ g/cm$^3$ and $T \approx 1$ GPa are the mass density and tensile stress of Si$_3$N$_4$. To increase the overlap between the optical field and the membrane, a third-order TM-polarized optical whispering gallery mode is specifically chosen, with the mode profile shown in the inset (b) in Fig. 1.

The dielectric membrane in the evanescent field of the microcavity would in principle give rise to both dispersive (frequency shift of resonance) and reactive (extra dissipation) contributions to the optical-cavity response [20]. According to the first order perturbation theory, the frequency-shift $\Delta\omega = \omega - \omega_0$ induced by the dielectric membrane can be written as [6]:

$$\Delta\omega = -\frac{\omega_0}{2} \frac{\int_{V_{\text{mem}}} (\varepsilon(\vec{r})-1)|E(\vec{r})|^2 \, d\vec{r}}{\int_V \varepsilon(\vec{r})|E(\vec{r})|^2 \, d\vec{r}}, \tag{1}$$

where $\omega_0$ is the intrinsic cavity frequency, $\varepsilon$ denotes the relative permittivity, $V_{\text{mem}}$ represents the volume of the membrane and $V$ stands for the whole space. We first consider the static case when the membrane is uniformly placed at a distance $z_0$ above the microtoroid (the inset in Fig. 2(b)). In the microtoroid, the optical field evanescently above the cavity can be approximated as $|E(x, y, z)|^2 \approx |E_0(x, y)|^2 e^{-z/L}$, where $E_0(x, y)$ is the electric field strength at $z = 0$ plane (the inset in Fig. 2(b)), $L$ is the decay length of the evanescent field. Therefore, the frequency shift $\Delta\omega_s(z_0) = \omega_s(z_0) - \omega_0$ in the static case can be written as

$$\Delta\omega_s(z_0) = -\frac{\omega_0 L A_{\text{int}}}{2V_{\text{cav}}}(1-e^{-h/L})(\varepsilon_{\text{SiN}}-1)\left|\frac{E_{0m}}{E_m}\right|^2 e^{-z_0/L}, \tag{2a}$$

$$A_{\text{int}} = \frac{1}{|E_{0m}|^2} \int_{-l_x/2}^{l_x/2} dx \int_{-l_y/2}^{l_y/2} dy |E_0(x, y)|^2, \tag{2b}$$

$$V_{\text{cav}} = \frac{1}{|E_m|^2} \int_V \varepsilon(\vec{r})|E(\vec{r})|^2 \, d\vec{r}. \tag{2c}$$

Here, $A_{\text{int}}$ describes the effective interaction area between the membrane and the optical mode, $V_{\text{cav}}$ is the optical mode volume, $E_{0m}$ and $E_m$ denote the maximum electric field strength at $z = 0$ plane and inside the microtoroid, respectively. In the system, the evanescent field above the cavity mainly distributes upon the microcavity perimeter so that the frequency shift induced by the membrane in the static case does not depend on $l_x$ and $l_y$ when $l_x, l_y > D$. Finite element simulation results of the frequency shift with $l_x, l_y > D$ are shown in Fig. 2(a). On one hand, the dispersive optomechanical interaction decay exponentially when $z_0$ increases as predicted by Eq. (2a), because of the evanescent nature of the optical field. On the other hand, the optomechanical interaction enhances as $d$ decreases from 3.0 μm to 2.4 μm. The enhanced interaction in the case of small $d$ result from two aspects: the reduced optical mode volume and the increased evanescent field strength located at the membrane. In the proposed system, although the relative field strength $|E_{0m}/E_m| \sim 0.15$ is several times smaller than that ($\sim 0.5$) in the toroid-string structure, the increased interaction area brings significantly enhanced optomechanical interaction. For instance, when $d = 2.4$ μm and $z_0 = 15$ nm (point A in Fig. 2(a)), the static linear and quadratic dispersive optomechanical coupling strengths reach $\omega_s' = d\omega_s/dz_0 = 2\pi \times 7$ GHz/nm and $\omega_s'' = d^2\omega_s/dz_0^2 = -2\pi \times 120$ MHz/nm$^2$. The reactive optomechanical interaction in the static case is simulated in Fig. 2(b) and can be analyzed in similar ways. Compared with the dispersive interaction, the reactive interaction is very weak.

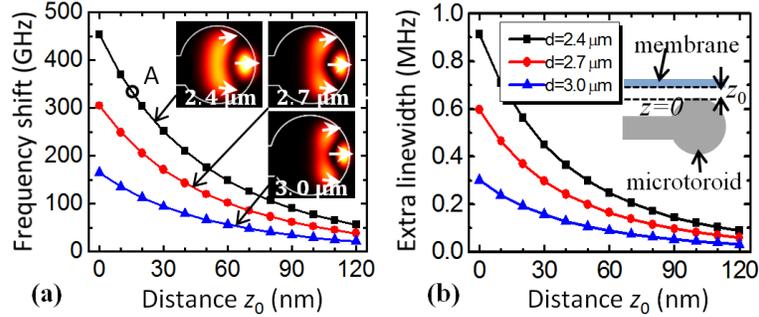

FIG. 2 (color online) Resonant frequency shift $|\Delta\omega_s/2\pi|$ (a) and the extra cavity linewidth $\Delta\kappa_s/2\pi$ (b) induced by the membrane in the static case. Black, red and blue curves correspond to $d = 2.4$ μm, 2.7 μm and 3.0 μm. Here, we fix $D = 20$ μm, $h = 50$ nm, $\lambda = 850$ nm, $n_{SiN} = 2.0 + 0.6 \times 10^{-6} i$ [21]. Insets of (a): Optical mode patterns in microtoroids with different minor diameters. Inset of (b): Illustration of the distance $z_0$ and the $z = 0$ plane.

Now we turn to investigate the dynamic case by considering the motion of the membrane. Simply speaking, the membrane mechanical motion $U_{j,k}(x, y, t) = \xi(t)u_{j,k}(x, y)$, with the motion amplitude $\xi(t)$ and the mechanical mode pattern $u_{j,k}(x, y) = \sin[j\pi(x/l_x+1/2)]\sin[k\pi(y/l_y+1/2)]$, will transduce to the frequency shift and linewidth variation of the cavity. According to Eq. (1), the frequency-shift $\Delta\omega_d(z_0) = \omega_d(z_0) - \omega_0$ dynamically induced by the membrane can be written as:

$$\Delta\omega_d = -B \int_{-l_y/2}^{+l_y/2} dy \int_{-l_x/2}^{+l_x/2} dx \int_{z_0}^{z_0+h} dz \left|E(x, y, z+U_{j,k})\right|^2, \qquad (3)$$

where $B = \omega_0(\varepsilon_{SiN}-1)/(2V_{cav}|E_m|^2)$. By expanding Eq. (3) to the second order of $\xi(t)$, we obtain

$$\Delta\omega_d = \Delta\omega_s + \omega_d'\xi + \frac{1}{2}\omega_d''\xi^2 + o(\xi^3), \qquad (4a)$$

$$\omega_{\text{d}}' = -B \int_{-l_y/2}^{+l_y/2} dy \int_{-l_x/2}^{+l_x/2} dx \int_{z_0}^{z_0+h} dz \cdot u_{j,k}(x,y) \frac{\partial}{\partial z} |E(x,y,z)|^2, \quad (4b)$$

$$\omega_{\text{d}}'' = -B \int_{-l_y/2}^{+l_y/2} dy \int_{-l_x/2}^{+l_x/2} dx \int_{z_0}^{z_0+h} dz \cdot u_{j,k}^2(x,y) \frac{\partial^2}{\partial z^2} |E(x,y,z)|^2, \quad (4c)$$

where $\omega_{\text{d}}'$ and $\omega_{\text{d}}''$ are dynamic linear and quadratic dispersive coupling rates, respectively. In the system, the dynamic reactive coupling can also be analyzed as above. Since the optical linewidth variation induced by the motion of the membrane is much smaller than both the dynamic frequency shift and the total optical linewidth, the dynamic reactive coupling can be neglected. Therefore, the optomechanical interaction is described by the dispersive Hamiltonian $H_{\text{int}} = \hbar \hat{a}^+ \hat{a} \left[ g^{(1)}(\hat{b}^+ + \hat{b}) - (1/2) g^{(2)}(\hat{b}^+ + \hat{b})^2 \right]$, where $\hat{a}$ and $\hat{b}$ are annihilation operators of the optical and mechanical modes. The single-photon linear and quadratic optomechanical coupling rates are given by $g^{(1)} = x_{\text{zpf}} \omega_{\text{d}}'$ and $g^{(2)} = -x_{\text{zpf}}^2 \omega_{\text{d}}''$, where $x_{\text{zpf}} = \sqrt{\hbar / 2 m_{\text{eff}} \omega_{\text{m}}}$ is the mechanical zero-point amplitude and the effective mass $m_{\text{eff}}$ is 1/4 of the physical mass of the membrane for all mechanical modes.

In the proposed system, the mechanical modes have different symmetrical properties viewed from the microcavity, so that they can interact either linearly or quadratically with the rotational symmetrical optical field. Generally, when the antinode of the membrane mode is above the center of the toroid, as shown in Fig. 3(a) for the mode with $(j, k) = (1, 1)$, the quadratic interaction can be neglected and the coupling is linear, like most optomechanical systems. On the contrary, if the node of the membrane mode is just above the center of the toroid, $g^{(1)}$ would vanish, hence the optomechanical coupling becomes purely quadratic, as shown in Fig. 3(b) for the mode with $(j, k) = (1, 2)$.

Next we estimate the single-photon coupling rates by considering the properties of the evanescent field. First, with the approximation $|E(x, y, z)|^2 \approx |E_0(x, y)|^2 e^{-z/L}$, Eqs. 4(b) and 4(c) can be written as $\omega_{\text{d}}' = \eta_{j,k}^{(1)} \omega_{\text{s}}'$ and $\omega_{\text{d}}'' = \eta_{j,k}^{(2)} \omega_{\text{s}}''$, respectively, where $\eta_{j,k}^{(1)}$ and $\eta_{j,k}^{(2)}$ are coefficients that bridge the dynamic and static optomechanical coupling rates ($m = 1, 2$)

$$\eta_{j,k}^{(m)} = \frac{1}{A_{\text{int}} |E_{0m}|^2} \int_{-l_y/2}^{+l_y/2} dy \int_{-l_x/2}^{+l_x/2} dx |E_0(x,y)|^2 u_{j,k}^m(x,y). \quad (5)$$

Then, considering that the evanescent field above the cavity mainly distributes upon the microcavity perimeter and that $l_x$, $l_y > D \gg d$, coefficients $\eta^{(1)}_{j,k}$ and $\eta^{(2)}_{j,k}$ can be approximated as follows ($m = 1, 2$)

$$\eta^{(m)}_{j,k} = \frac{1}{2\pi}\int_0^{2\pi} u^m_{j,k}(\frac{D}{2}\cos\theta, \frac{D}{2}\sin\theta)\mathrm{d}\theta. \qquad (6)$$

Now with $g^{(1)} = \eta^{(1)}_{j,k} x_{\mathrm{zpf}} \omega_s'$ and $g^{(2)} = -\eta^{(2)}_{j,k} x^2_{\mathrm{zpf}} \omega_s''$ we are able to investigate the single-photon coupling rates numerically. The dependence of $g^{(1)}$ on the size of the membrane for $(j, k) = (1, 1)$ is plotted in Fig. 3(c). It can be seen that the system achieves one-order-of-magnitude increased $g^{(1)}/2\pi$ (> 20 kHz) comparing with the near-field toroid-string optomechanical systems ($g^{(1)}/2\pi \sim$ 1 kHz). Moreover, since the system enables simultaneously low mechanical and optical quality losses, the single-photon cooperativity $C = 4(g^{(1)})^2/\kappa\gamma_m$ reaches 320, with $\{j, k, l_x, l_y, \omega_m/2\pi, \kappa/2\pi, \gamma_m/2\pi\} = \{1, 1, 40\ \mu\mathrm{m}, 40\ \mu\mathrm{m}, 10\ \mathrm{MHz}, 5\ \mathrm{MHz}, 1\ \mathrm{Hz}\}$. The variation of $g^{(2)}$ as a function of the size of the membrane for $(j, k) = (1, 2)$ is depicted in Fig. 3(d). It is found that our system achieves a much larger $g^{(2)}/2\pi$ (> 0.7 mHz) compared with previous work [22]. Here, we note that both $g^{(1)}$ and $g^{(2)}$ can be maximized in the system by optimizing the size of the membrane. On one hand, a smaller membrane provides a larger mechanical zero-point amplitude $x_{\mathrm{zpf}}$ because it has a smaller mass. On the other hand, both $\eta^{(1)}_{j,k}$ and $\eta^{(2)}_{j,k}$ can be controlled since they strongly depend on the local amplitude of the mechanical mode upon the microcavity perimeter, where the evanescent optical fields above the cavity mainly distribute.

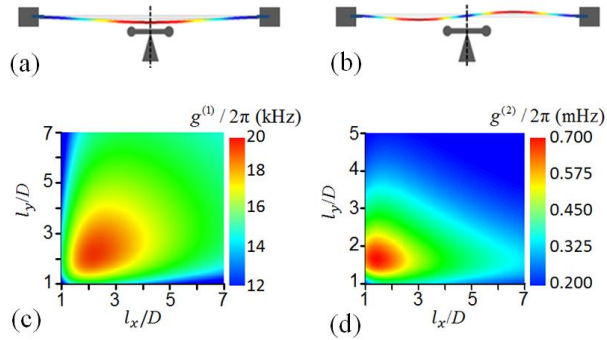

FIG. 3 (color online) (a), (b) Cross sectional views ($x=0$) for exemplary linear (a) and quadratic (b) coupling conditions, corresponding to membrane modes $(j, k) = (1, 1)$ and $(1, 2)$. (c) Linear vaccum optomechanical coupling rate $g^{(1)}$ for mechanical mode $(j, k) = (1, 1)$. (d) Quadratic single-photon optomechanical coupling rate $|g^{(2)}|$ for mechanical mode $(j, k) = (1, 2)$. In both (c) and (d), $D = 20\ \mu\mathrm{m}$, $d = 2.4\ \mu\mathrm{m}$, $h = 50$ nm, and $z_0 = 15$ nm.

In the system, the purely quadratic coupling relies on the membrane node being exactly above the microtoroid and the membrane being perfectly horizontal relative to the microtoroid. Experimentally, deviations from that will introduce linear optomechanical coupling. Therefore, it is important to discuss the requirements for investigating quadratic coupling phenomena. Generally, to investigate quadratic coupling phenomena, optomechanical systems are driven on resonance with the optical mode or are driven on the red (or blue) two-phonon resonance [15, 23]. In the resolved sideband limit $\omega_m \gg \kappa$, the condition $|g^{(1)}/g^{(2)}| < 1$ enables us to neglect the off-resonant terms of the linear optomechanical coupling with the rotating wave approximation and to concentrate on the effects of the quadratic coupling. For the proposed system, we consider the case in which the center of the membrane deviates $\Delta x$ and $\Delta y$ along $x$ and $y$ axis respectively and the membrane tilts $\alpha_x$ and $\alpha_y$ around $x$ and $y$ axis respectively. With small displacements $\Delta x$, $\Delta y \ll l_x$, $l_y$ and tilting angles $\alpha_x$, $\alpha_y \ll 1$, the motion of the mechanical membrane can be expressed as $U_{j,k}(x, y, t) = \xi(t)u_{j,k}(x', y') + (\alpha_x y' + \alpha_y x')$, where $x' = x - \Delta x$ and $y' = y - \Delta y$. By substituting $U_{j,k}(x, y, t)$ into Eq. (3) and expand it to the second order of $\xi(t)$ we derive the coefficients $\eta_{j,k}^{(1)}$ and $\eta_{j,k}^{(2)}$ with the approximations applied in deriving Eq. (6) ($m = 1, 2$)

$$\eta_{j,k}^{(m)} = \frac{1}{2\pi} \int_0^{2\pi} u_{j,k}^m(x'(\theta), y'(\theta)) e^{-[\alpha_x y'(\theta) + \alpha_y x'(\theta)]/L} \, d\theta, \tag{7}$$

where $x'(\theta) = (D/2)\cos\theta - \Delta x$ and $y'(\theta) = (D/2)\sin\theta - \Delta y$. For $\{j, k, l_x, l_y\} = \{1, 2, 40\ \mu m, 40\ \mu m\}$, we investigate the ratio $|g^{(1)}/g^{(2)}|$ in consideration of the displacements and the tilting angles in Figs. 4(a) and 4(b), respectively. It can be seen that the ratio $|g^{(1)}/g^{(2)}|$ is insignificantly effected by $\Delta x$ and $\alpha_y$. Thus, experimentally we are allowed to focus on $\Delta y$ and $\alpha_x$. We emphasize here that the weak dependence of $|g^{(1)}/g^{(2)}|$ on $\Delta x$ and $\alpha_y$ is a result of choosing the mode $(j, k) = (1, 2)$, and that for $(j, k) = (2, 1)$, the roles of $\alpha_x$ and $\alpha_y$ ($\Delta x$ and $\Delta y$) interchange. For $(j, k) = (1, 2)$, the condition $|g^{(1)}/g^{(2)}| < 1$ is satisfied with $|\Delta y| < 0.5$ pm and $|\alpha_x| < 0.3$ nrad. The displacement requirement (0.5 pm) is as large as the displacement suggested in Ref. 15. Since the required tilting angle (0.3 nrad) of the membrane is far from the tilting angle achieved experimentally in the membrane-in-the-middle setups ($\sim 10^4$ nrad) [24], investigations of quadratic optomechanical interaction with the proposed system demand experimental improvement of controlling the tilting angles of the membrane.

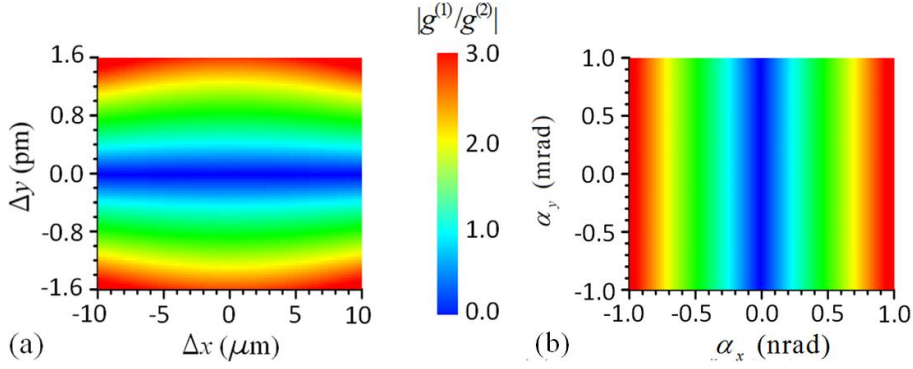

FIG. 4 (color online) (a) The ratio $|g^{(1)}/g^{(2)}|$ as a function of $\Delta x$ and $\Delta y$, with $\alpha_x = \alpha_y = 0$. (b) The dependence of $|g^{(1)}/g^{(2)}|$ on $\alpha_x$ and $\alpha_y$, with $\Delta x = \Delta y = 0$. Here, $\{D, d, l_x, l_y, h, z_0, j, k\} = \{20\ \mu m, 2.4\ \mu m, 40\ \mu m, 40\ \mu m, 50\ nm, 15\ nm, 1, 2\}$.

Benefiting from the ultrahigh single-photon cooperativity, the system is promising to realize optical nonlinearity at low light intensities, which is of great interest for both fundamental and applied studies. For example, the strong photon-photon interaction can be used for low-threshold all-optical switching. Generally, with a control pulse driving at the red sideband, an optomechanical system enables the phenomenon analogy to electromagnetically induced transparency [9, 10, 25] in the regime $\omega_m > \kappa > \sqrt{n}g^{(1)} > \gamma_m$. As depicted in the inset of Fig. 5(a), an incoming probe pulse can be switched from reflecting to transmitting via the control pulse. In the critical taper-cavity coupling condition, the peak value of the transparency window is given by $(nC)^2/(nC+1)^2$. Therefore, the efficient optomechanical switching essentially requires $n>1/C$. So far, optomechanical switching has been realized with five intracavity control photons, since the single-photon cooperativity $C$ is only about 0.2 [10]. Remarkably, $C$ reaches 320 in the present optomechanical system, and as shown in Fig. 5(a), the switching can be implemented at the level of 0.01 intracavity control photons, corresponding to an ultralow input power of tens of fW.

The optomechanical switching can be investigated more specifically. On the condition $n\gg1/C$, the resonant probe pulse experiences a group delay of $\tau_d = 2/(nC\gamma_m)$. This indicates that the switching process must last longer than $\tau_d$ for a given intracavity control photon number $n$, or equally for a given control pulse power $P$ with the standard coupled mode theory. Thus, we derive that an input control pulse carrying more than $(\omega_m^2+\kappa^2/4)/(g^{(1)})^2$ photons is demanded for efficient optomechanical switching, where the $n$ is cancelled in the deduction. Considering that $\omega_m > \kappa$ is necessary for the switching process, the photon number $N_c \sim (\kappa/g^{(1)})^2$ can be defined as the switching threshold for the input control pulse in the limit $\omega_m \sim \kappa$. Since our system supports both large linear optomechanical coupling rate $g^{(1)}$ and small optical decay rate $\kappa$, a switching threshold $N_c$ as low as $10^4$ in optomechanics is accessible as shown in Fig. 5(b). This control photon number threshold is unprecedented in optomechanical systems and is comparable to many other all-optical switches [26]. In addition, compared with other optical switches by using atomic ensembles [27] and cavity quantum electrodynamics [28], solid state optomechanical switches are easy to be integrated on chips and operated in a large and technologically relevant range of frequencies.

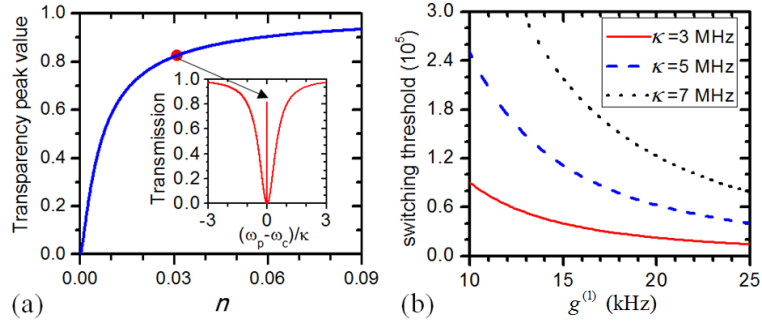

FIG. 5 (color online) (a) Transparency peak value against intracavity photon number $n$ under the critical taper-cavity coupling condition. Here, $\{j, k, l_x, l_y, \omega_m/2\pi, g^{(1)}/2\pi, \kappa/2\pi, \gamma_m/2\pi\} = \{1, 2, 40\ \text{mm}\ 40\ \text{mm}, 10\ \text{MHz}, 20\ \text{kHz}, 5\text{MHz}, 1\text{Hz}\}$. Inset: Transmission spectrum of the probe field with $n = 0.03$. (b) The optomechanical switching threshold $N_c$ as a function of $g^{(1)}$ for $\kappa/2\pi = 3, 5, 7$ MHz.

Combining the enhanced coupling rate and high mechanical quality factor is also crucial for overcoming the thermal decoherence in quantum implementation, such as achieving high temperature entanglement between the optical field and the mechanical oscillators [29, 30]. We calculate the logarithmic negativity $E_N$ (as a measure of the entanglement [31]) with parameters of the proposed system $\{\omega_m/2\pi, g^{(1)}/2\pi, \kappa/2\pi\} = \{10\ \text{MHz}, 20\ \text{kHz}, 30\ \text{MHz}\}$. For an accessible mechanical quality $Q_m = 10^6$ at high temperature and a low input power 5 μW driving at the red sideband, the resulted effective coupling rate is strong enough to overcome the thermal decoherence. As a result, a robust optomechanical entanglement can be obtained and $E_N$ persists higher than 0.05 even at room temperature. More quantum phenomena can be investigated and applied under the strong coupling regime ($\sqrt{n}g^{(1)} > \kappa, \Gamma_{th}$) [32, 33], which can be achieved in the system under experimentally accessible condition even at room temperature.

The enhanced quadratic coupling in the optomechanical system is desired for the quantum non-demolition (QND) measurement of mechanical energy, such as the phonon shot noise [19, 34]. In the resolved sideband limit $\omega_m \gg \kappa$, the quadratic optomechanical interaction Hamiltonian of the system can be approximated as $H_{int} = -\hbar g^{(2)}(\hat{b}^+\hat{b}+1/2)\hat{a}^+\hat{a}$. This indicates that a change of the membrane phonon number would lead to a resonance frequency shift of the optical mode, and the shift is strongly dependent on the quadratic single-photon optomechanical coupling strength. In our system, $g^{(2)}/2\pi$ is significantly increased, reaching 0.7 mHz. As a result, the signal-to-noise ratio for the QND measurement of phonon shot noise is over 100 with other parameters used in Ref. [19], as listed in Table 1. In addition, the enhanced quadratic coupling also enables other applications such as two-phonon laser cooling and optical squeezing [23].

TABLE 1. Parameters allowing a signal-to-noise ratio $S = 102$ for the phonon shot noise detection in our system. The membrane is laser cooled to $n_m = 0.2$ (effective thermal quanta) from a starting temperature $T = 300$ mK and is driven to 1 nm amplitude of motion. The input power is 5 µW at 850 nm for detection.

| $l_x \times l_y \times h$ | $j, k$ | $\omega_m/2\pi$ | $Q_m$ | $\kappa/2\pi$ | $g^{(2)}/2\pi$ |
|---|---|---|---|---|---|
| 50 µm × 50 µm × 50 nm | 1, 2 | 12 MHz | $1.2 \times 10^7$ | 5 MHz | 0.5 mHz |

In conclusion, we have studied a near-field optomechanical system consisting of a high-mechanical-quality silicon nitride membrane and a high-optical-quality silica microtoroid. The increased interaction area between them leads to significantly enhanced linear and quadratic optomechanical interaction. Benefiting from large single-photon optomechanical coupling rates ($g^{(1)}$ and $g^{(2)}$) and simultaneously high optical and mechanical quality factors, the system enables studies of optical nonlinearity at low light intensities and phonon shot noise of mechanical oscillators. Due to the advantage of overcoming thermal decoherence, it holds great potential in realizing optomechanical entanglement at room temperature, quantum state transfer between single optical photons and mechanical phonons [35, 36], quantum transducers [37] and quantum memory elements [38, 39].

This work was supported by the NSFC (Grants No. 11004003 and No. 11121091) and the Research Fund for the Doctoral Program of Higher Education (No. 2009001120004). H.K.L, X.Y. and X.X.R. were supported by the National Fund for Fostering Talents of Basic Science (Nos. J1030310 and J1103205), the Chun-Tsung Scholar Fund for Undergraduate Research of Peking University, and the Undergraduate Research Fund of Education Foundation of Peking University. H.K.L. and Y.C.L. contributed equally.